\documentclass[]{aastex631}
\usepackage{graphicx}
\usepackage{verbatim}
\usepackage{color}
\usepackage{natbib}
\usepackage{booktabs}
\usepackage{tabularx}
\usepackage{amsmath}
\usepackage{array}
\usepackage{multirow}
\usepackage{makecell}
\usepackage[normalem]{ulem}
\usepackage{xspace}
\usepackage{xcolor}
\usepackage[varg]{txfonts}
\usepackage{colortbl,dcolumn}
\usepackage{tabularx}
\usepackage{hyperref}

\newcommand{\SubItem}[1]{
    {\setlength\itemindent{18pt} \item[*] #1}
}

\shorttitle{Synthetic image generation}
\shortauthors{Salvatelli et al.}
\graphicspath{{./}{figures/}}

\begin{document}

\title{Exploring the Limits of Synthetic Creation of Solar EUV Images via Image-to-Image Translation}

\correspondingauthor{Valentina Salvatelli}
\email{vsalvatelli@microsoft.com}

\author[0000-0002-3232-4101]{Valentina Salvatelli}
\affiliation{Microsoft Research, Cambridge CB12FB, UK}
\affiliation{Frontier Development Lab, Mountain View, CA 94043, USA}
\affiliation{SETI Institute, Mountain View, CA 94043, USA}

\author[0000-0001-5190-442X]{Luiz F. G. dos Santos}
\affiliation{Shell Global Solutions International B.V., Grasweg 31, 1031 HW Amsterdam, The Netherlands}
\affiliation{nextSource Inc, New York, NY 10018, USA}

\author[0000-0002-2180-1013]{Souvik Bose}
\affiliation{Rosseland Center for Solar Physics, University of Oslo,P.O. Box 1029 Blindern, NO-0315 Oslo, Norway}
\affiliation{Institute of Theoretical Astrophysics, University of Oslo,P.O. Box 1029 Blindern, NO-0315 Oslo, Norway}
\affiliation{Lockheed Martin Solar \& Astrophysics Laboratory, Palo Alto, CA 94304, USA}
\affiliation{Bay Area Environmental Research Institute, NASA Research Park, Moffett Field, CA 94035, USA}

\author{Brad Neuberg}
\affiliation{Frontier Development Lab, Mountain View, CA 94043, USA}
\affiliation{SETI Institute, Mountain View, CA 94043, USA}
\affiliation{Planet, San Francisco, CA 94107, USA}

\author[0000-0003-2110-9753]{Mark C. M. Cheung}
\affiliation{Lockheed Martin Solar \& Astrophysics Laboratory, Palo Alto, CA 94304, USA}

\author[0000-0002-6203-5239]{Miho Janvier}
\affiliation{Universit\'e Paris-Saclay, CNRS, Institut d'astrophysique spatiale, Orsay, France}

\author[0000-0002-9672-3873]{Meng Jin}
\affiliation{SETI Institute, Mountain View, CA 94043, USA}
\affiliation{Lockheed Martin Solar \& Astrophysics Laboratory, Palo Alto, CA 94304, USA}

\author[0000-0002-2733-2078]{Yarin Gal}
\affiliation{OATML Group, Department of Computer Science, University of Oxford, UK}

\author[0000-0001-9854-8100]{At{\i}l{\i}m G\"{u}ne\c{s} Bayd{\rlap{\.}\i}n}
\affiliation{Department of Computer Science, University of Oxford, Oxford OX1 3QD, UK}



\begin{abstract}
The Solar Dynamics Observatory (SDO), a NASA multi-spectral decade-long mission that has been daily producing terabytes of observational data from the Sun, has been recently used as a use-case to demonstrate the potential of machine learning methodologies and to pave the way for future deep-space mission planning. In particular, the idea of using image-to-image translation to virtually produce extreme ultra-violet channels has been proposed in several recent studies, as a way to both enhance missions with less available channels and to alleviate the challenges due to the low downlink rate in deep space. This paper investigates the potential and the limitations of such a deep learning approach by focusing on the permutation of four channels and an encoder--decoder based architecture, with particular attention to how morphological traits and brightness of the solar surface affect the neural network predictions. In this work we want to answer the question: can synthetic images of the solar corona produced via image-to-image translation be used for scientific studies of the Sun? The analysis highlights that the neural network produces high-quality images over three orders of magnitude in count rate (pixel intensity) and can generally reproduce the covariance across channels within a 1\% error. However the model performance drastically diminishes in correspondence of extremely high energetic events like flares, and we argue that the reason is related to the rareness of such events posing a challenge to model training.

\end{abstract}




\section{Introduction}

Since its launch in 2010, NASA's Solar Dynamics Observatory~\citep[SDO;][]{SDO_primary} has monitored the evolution of the Sun. SDO data has enabled researchers to track the evolution of the Sun's interior plasma flows over solar cycle 24 and beyond. It has also continuously monitored the evolution of the solar corona, capturing dynamical evolution at time-scales of seconds and minutes. This capability is due to the suite of four telescopes on the Atmospheric Imaging Assembly~\citep[AIA;][]{AIA} instrument, which captures full-Sun images at two ultraviolet (UV) bands, seven extreme UV (EUV) bands, and one visible band. The seven EUV channels are designed to capture photons from emission lines in highly ionized metals in plasmas at transition region (TR; $10^5$ K $\lesssim T \lesssim 10^6$ K) and coronal temperatures ($10\gtrsim 10^6$ K). This combination of channels with sensitivity to different temperatures allows researchers to track how transition regions and coronal plasmas heat and cool~\citep[e.g.,][]{Cheung:2015}, and to use these thermal histories to test theories of coronal heating and of flares. 

The high spatial resolution ($\bf \sim 1.5\arcsec$, $4096\times 4096$ pixels), high cadence (12 s for EUV channels) full-disk observing capability is possible because of SDO's ground system providing a sustained downlink rate of $\sim~67$ Mbps. The collection of continuous data, over more than one solar cycle, provides not only numerous opportunities to perform data-driven scientific studies but also research with the potential to help optimize future solar physics missions. 

For instance, the idea of using SDO images for image-to-image translation has been explored in several papers, most notably by \cite{Baso2018, SDOML, Szenicereaaw6548, Park_2019, salvatelli2019}. Image-to-image translation can potentially provide a way to enhance the capabilities of solar telescopes with fewer channels or less telemetry than is available to SDO. The \emph{SDO image translation problem} can be defined as follows: given a set of $N$ (nearly) contemporaneous images taken in different EUV channels, can a model be developed which maps the $N$ input images to the image of a missing (not in input) EUV channel?

Notably, \cite{Lim_2021} adopted a widely used image translation method \citep[Pix2Pix,][]{pix2pix} to tackle the SDO image-translation problem and to understand which subset of channels can better translate other channels. They trained and evaluated models for all combinations of input channels for both $N=2$ and $N=3$ variants of the problem, and compared global image quality metrics to pick out the channel combinations that perform the best. For some channel combinations, the reported pixel-to-pixel correlation coefficient approaches unity. 

In this paper, we build on the method presented in \cite{salvatelli2019} for one single channel and we delve deeper into the opportunities and the limitations of applicability of such ``virtual telescopes''. We focus on a permutation of a subset of channels (4 out of 10) and we explore in greater detail what is the quality of this synthetic generation on a number of scientifically-motivated metrics (figures of merit) and in relation to periods and regions of different level of activity of the Sun.

Together with this paper we also open source the code we used for the analysis~\cite{salvatelli_valentina_2022_6954828}\footnote{\href{https://zenodo.org/record/6954828.YumocezMJ-U}{Zenodo: ML pipeline for Solar Dynamics Observatory (SDO) data}} and that can be used by the community to train  and evaluate similar models on the publicly available SDO dataset released by \cite{SDOML} .

\section{Data}
\label{sec.data}
The work presented in this project is based on data from SDO's AIA. The AIA instrument takes full-disk, 
$4096 \times 4096$ pixel, imaging observations of the solar photosphere, chromosphere and corona in two UV channels and in seven extreme UV (EUV) channels. The original SDO dataset was processed in \cite{SDOML} into a machine-learning ready dataset of $\sim6.6$ TB (hereafter SDOML) that we leveraged for the current work. 

The SDOML dataset is a subset of the original SDO data ranging from 2010 to 2018. Images are spatially co-registered, have identical angular resolutions, are corrected for the instrumental degradation over time and have exposure corrections applied. All the instruments are temporally aligned. AIA images in the SDOML dataset are available at a sampling rate of 6 min. The $512\times512$ pixel full-disk images have a pixel size of $\thicksim 4\farcs8$. 

The images are saved in single-precision floating point to preserve the high dynamic range ($\gtrsim 14$ bits per channel per pixel). For numerical performance purposes, the images of each channel are re-scaled by a per-channel constant factor which is approximately the average count rate for that channel. The per-channel constant factors can be found at Tab.\ref{tab:average_channels}.

\section{Methodology} 
\label{sec.meth}
Our approach of synthesizing solar EUV images is to perform image translation from multiple input channels to one single output channel. For the development of this work we focused on the permutations of four channels (94, 171, 193, 211 \AA). These channels are sensitive to coronal plasmas at different temperatures~\citep{Cheung:2015}.

\begin{figure}[!htb]
  \centering
  \includegraphics[width=0.9\linewidth]{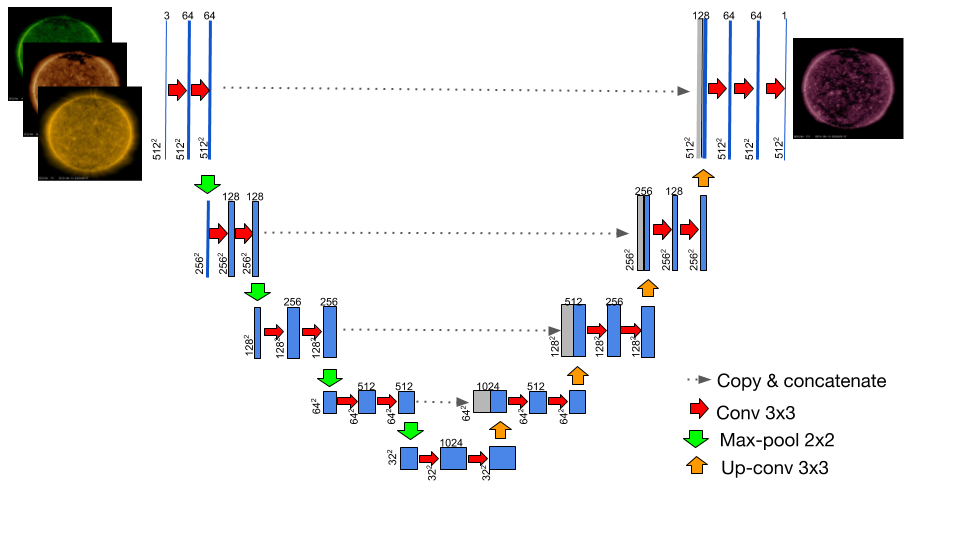}
  \caption{U-Net based architecture used to synthesize solar EUV images. Each box corresponds to a multi-channel feature map. Grey boxes are copied maps. The number of channels is shown on top of the box. Resolution in pixels is indicated on the left of the box. Arrows represent operations. For images of size $512\times512$, the trainable parameters are $34,513,857$. Figure taken from \citep{salvatelli2019}.}
  \label{fig.u-net}
\end{figure}

To perform the image translation we used a deep neural network \citep[DNN, ][]{goodfellow2016deep}, more specifically we adopted a U-Net architecture \citep{Unet15}, an encoder--decoder with skip connections that was first designed for image segmentation on medical images. We used Adam optimizer \citep{Optimizer} and Leaky ReLU~\citep{Maas13rectifiernonlinearities} activations, and implement the code using the open source library PyTorch \citep{paszke_2017}. The full details of the adopted architecture is given in Fig. \ref{fig.u-net}.  
We limit the number of channels to four for computational resources constraints. For the training and inference of the architecture presented above we used $4 \times$ NVIDIA Tesla T4s. We trained each model for 600 epochs.

For comparison we experimented also with a simpler baseline model, described by the following equation: 
\begin{equation}
\label{Eq.liner_model}
Y_{\rm pred} = \alpha X_{1} + \beta X_{2} + \gamma X_{3} + \delta
\end{equation}
where $Y_{\rm pred}$ is the reconstructed pixel of the output channel, $X_i$ are the pixel values of the input channels; $\alpha$, $\beta$, $\gamma$ are the weights and $\delta$ the bias of the linear combination of the channels. $\alpha$, $\beta$, $\gamma$, $\delta$ are trainable parameters of the model.

The metrics we use to evaluate the accuracy of our results for each permutation are:
\begin{itemize}
\item The difference between predicted and ground truth images in the form of normalized mean squared error (NMSE; Eq.~\ref{nmse}) and normalized root mean squared error (NRMSE; Eq.~\ref{rnmse}). \\
\begin{equation}
 {\rm NMSE}(y, \hat{y}) = \frac{\sum_{i=1}^{N}(y_i-\hat{y}_i)^2}{\sum_{i=1}^{N}{y_i}^2}
\label{nmse}
\end{equation} 
\begin{equation}
 {\rm RNMSE}(y, \hat{y}) = \frac{\sqrt{\frac{\sum_{i=1}^{N}(y_i-\hat{y}_i)^2}{N}}}{\overline{y}}
\label{rnmse}
\end{equation}
\item The structural similarity index \citep[SSIM;][]{Wang04imagequality}, a metric commonly used in computer vision to compute similarity between images, measuring the difference in terms of visually perceived texture and morphology. Identical images have SSIM equal to 1.
\item The average of NRMSE and SSIM, as described in Eq.~\ref{metric}. Lower values mean better performance in this metric. \\
\begin{equation}
 {\rm Err}(y, \hat{y}) = \frac{{\rm NRMSE}(y, \hat{y}) + \left[1-|{\rm SSIM}(y, \hat{y})|\right]}{2}
\label{metric}
\end{equation}
\item The average pixel-to-pixel Pearson correlation coefficient.
\end{itemize}

 In order to assess how much the DNN is able to learn the physical correlations between channels and to correctly reproduce them in the synthetic images, we also evaluate the difference between the real and the synthetic covariance of the channels. With the aim of better understanding the error, in addition to the standard covariance we compute the neighborhood covariance. In this case the output is a map of the same size of the input images where each value in the map corresponds to the covariance on a squared patch centered in the pixel and of size  $20\times20$ pixels as described in Eq.~\ref{eq.patch_covariance}. \\
\begin{equation}
\label{eq.patch_covariance}
    cov_{patch} = \frac{\sum_{i}^{N}[(y_{i}-\bar{y})(\hat{y}_{i}-\bar{\hat{y}})]}{N-1}
\end{equation}\\
where $N$ is the total number of pixels in the patch.

Each model has been trained on $6,444$ images ($1,611$ timestamps, one image per channel for each timestamp) in the intervals January $1^{st}~ 2011$ to July $31^{st}~ 2011$ and January $1^{st}~ 2012$ to July $31^{st}~ 2012$. For testing $2,668$ images ($667$ timestamps) have been used, taken in the intervals August $1^{st} 2011$ to October $31^{st} 2011$ and August $1^{st} 2012$ to October $31^{st}~ 2012$. Each timestamp is at least $6$1 hours apart from the closest ones. These time ranges have been selected to ensure we were testing on images significantly different from the training ones. Only timestamps for which all the channels of interest were available have been included in the above datasets.

\section{Experiments} 
\label{sec.exp}
For this analysis we trained eight models using the data and architecture described in Sec.~\ref{sec.data} and Sec.~\ref{sec.meth}, two models for each of the four channels permutation. For each channel permutation we trained (1) a model where the input data was scaled by a constant factor (cf. Tab.~\ref{tab:average_channels}) and (2) a model where the square root of the input data was taken, in addition to the constant scaling. The second scaling technique is to explore the impact of pixels with extreme ranges on the training. Each model has been evaluated by studying both the aggregated performance on the full test data and the performance on specific timestamps. Namely timestamps in the neighborhood of Valentine's Day flare (2011-2-15:1:50:00 UT) and in a quiet day of the same month (2011-02-10 00:00:00). The focus of these experiments is to evaluate the robustness of the image-to-image translation approaches in normal and extreme conditions of the Sun's activity. For comparison, we trained also four linear models, one model for each of the four channels permutation, using Eq.~\ref{Eq.liner_model}and input scaled by a constant factor.

\section{Results}
 \label{results}
 \begin{table}
 \begin{tabular}{lcc|cc|cc|cc}
\toprule
\textbf{Deep Neural Network} &  \textbf{211\_sqr} &       \textbf{211} &  \textbf{193\_sqr} &       \textbf{193} &  \textbf{171\_sqr} &       \textbf{171} &   \textbf{94\_sq}r &        \textbf{94} \\
\midrule
NMSE             &  0.010024 &  0.008748 &  0.013414 &  0.013015 &  0.015270 &  0.010151 &  0.009482 &  0.013643 \\
NRMSE            &  0.195127 &  0.182286 &  0.225717 &  0.222332 &  0.240829 &  0.196360 &  0.189773 &  0.227641 \\
$|$1 - SSIM $|$           &  0.040844 &  0.046189 &  0.022866 &  0.024522 &  0.030636 &  0.034892 &  0.114447 &  0.138455 \\
(NRMSE + $|$1 - SSIM$|$)/2 &  0.117985 &  0.114237 &  0.124292 &  0.123427 &  0.135732 &  0.115626 &  0.152110 &  0.183048 \\
\bottomrule
\end{tabular}
\caption{Performance of the DNN on different permutations of input/output channels in the set (94, 171, 193, 211 \AA{}) and for different scaling of the input data. In every column the input channels are all but the one indicated in the column name that corresponds to the output channel. Each value is the mean over the whole test dataset. For each metric in this table lower is better. For 94 \AA{} the similarity index is higher than for the others channels, this can be explained by the fact the average value in this channel is higher and the metric is affected by the absolute values. See Sec.~\ref{sec.meth} for explanation of the metrics.}
\label{Tab-permutation}
\end{table}
\begin{table}
\centering
\begin{tabular}{ccccc}
\toprule
\textbf{Deep Neural Network}              & \multicolumn{4}{c}{Model output} \\ \cline{2-5}
{Scaling}              &    211~\AA       &    193~\AA       &     171~\AA      &   94 ~\AA \\ 
\midrule
Non Root &  $0.994 \pm 0.004$ &  $0.991 \pm 0.006$ &  $0.993 \pm 0.003$ &  $0.991 \pm 0.003$ \\
Root     &  $0.993 \pm 0.004$ &  $0.996 \pm 0.004$ &  $0.990 \pm 0.005$ &  $0.994 \pm 0.004$ \\

\bottomrule
\end{tabular}
\caption{DNN model. Average Pearson correlation coefficient pixel-to-pixel, mean and standard deviation over the full test dataset for permutations of input/output channels in the set (94, 171, 193, 211 Å). For each channel combination the average Pearson correlation coefficient pixel-to-pixel was calculated for both trained models, with and without root scaling. The results observed are impressive and in all cases the performance is superior to $0.99$.}
\label{Tab-correlation}
\end{table}

\begin{table}
\centering
\begin{tabular}{lc|c|c|c}
\toprule
\textbf{Linear Model} &       \textbf{211} &       \textbf{193} &       \textbf{171} &        \textbf{094} \\
\midrule
NMSE                   &  0.749594 &  0.742833 &  0.741476 &   0.875264 \\
NRMSE                  &  1.687336 &  1.679708 &  1.678174 &   1.823300 \\
1 - SSIM               &  0.588910 &  0.441623 &  0.490644 &   0.976495 \\
(NRMSE + $|$1 - SSIM$|$)/2 &  1.138123 &  1.060665 &  1.084409 &   1.399897 \\
\bottomrule
\end{tabular}
\caption{For comparison with Tab.~\ref{Tab-permutation}, performance of the linear model on different permutations of input/output channels in the set (94, 171, 193, 211 \AA{}) for standard (no square root) scaling. The DNN consistently improves results of one order of magnitude in each of these metrics. The comparison demonstrates non-linear patterns between channels are important for a correct reconstruction of the images.}
\label{Tab-permutation-linear}
\end{table}


 \begin{figure}[htb!]
  \centering
  \includegraphics[trim=0 0 0 0, clip, width=0.48\linewidth]{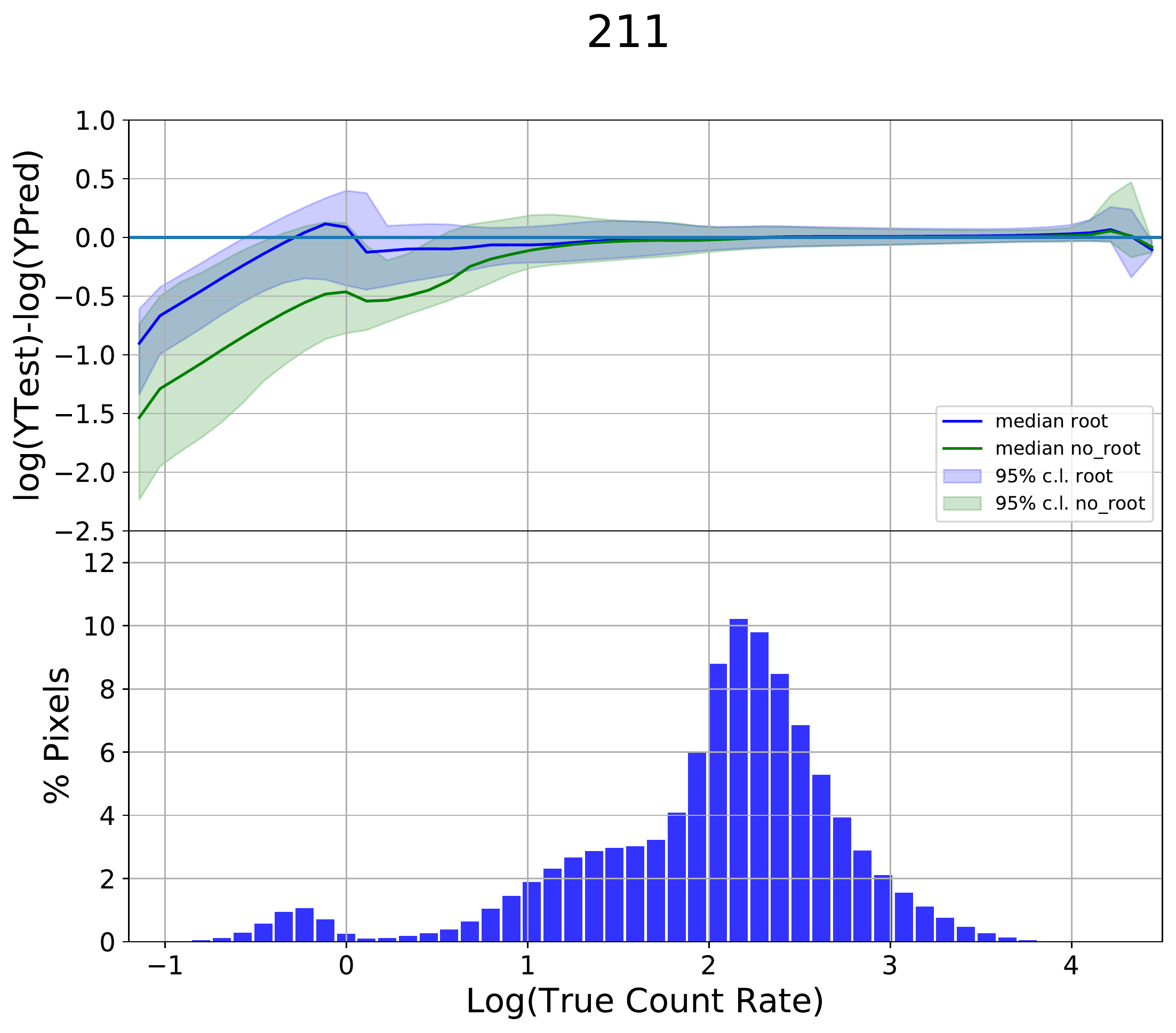}
  \includegraphics[trim=0 0 0 0, clip, width=0.48\linewidth]{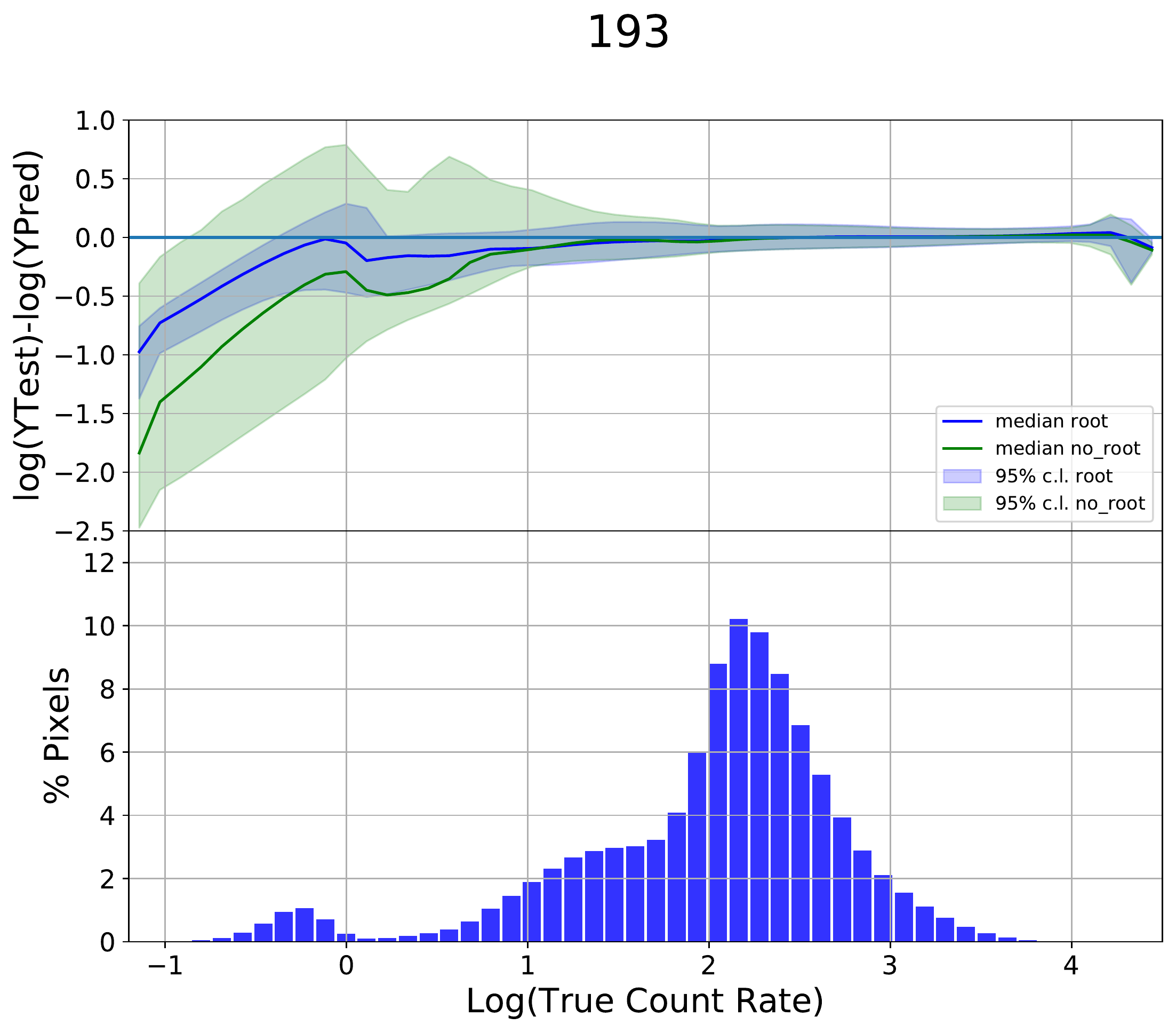}
  \newline
  \newline
  \includegraphics[trim=0 0 0 0, clip, width=0.48\linewidth]{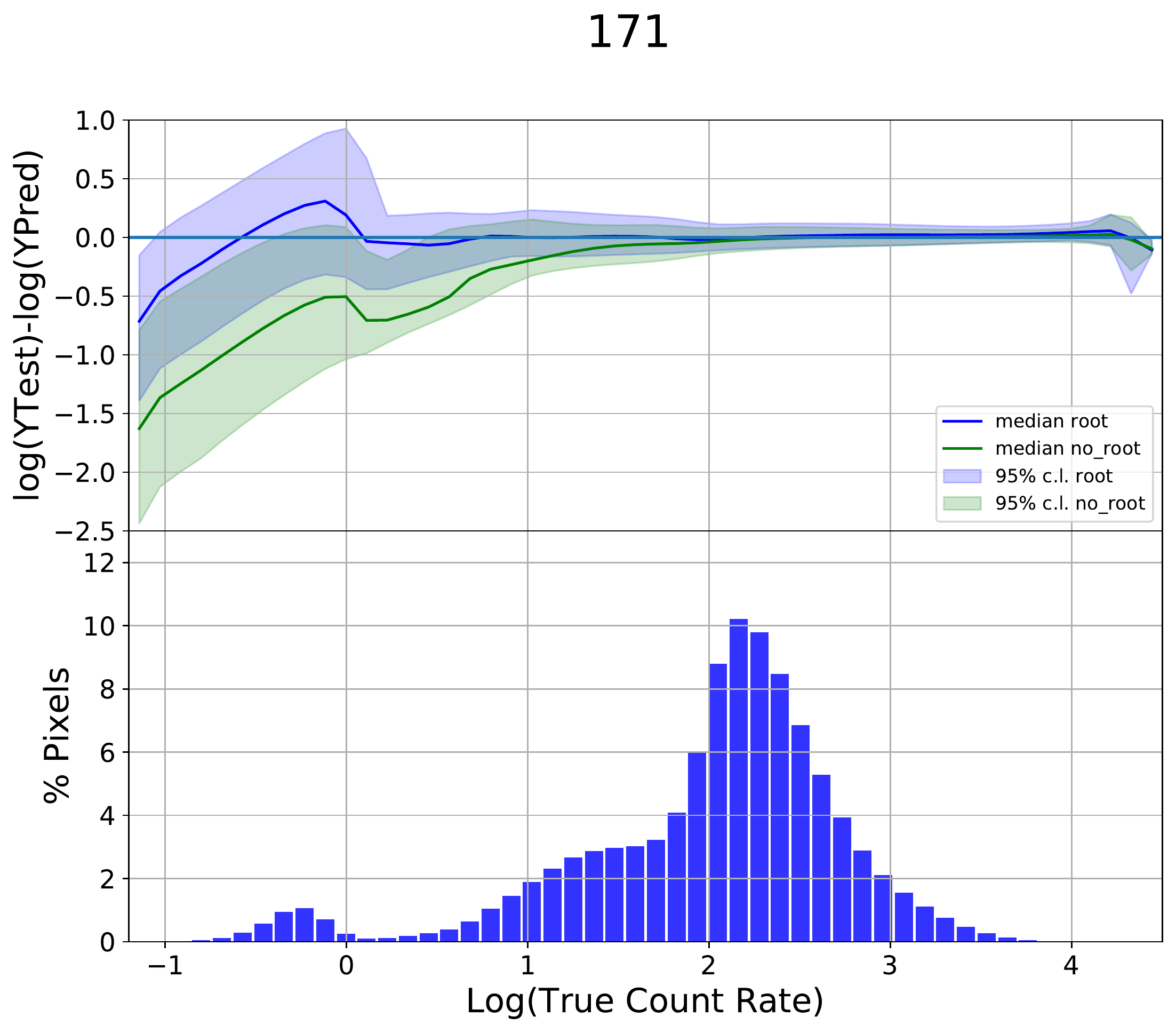}
  \includegraphics[trim=0 0 0 0, clip, width=0.48\linewidth]{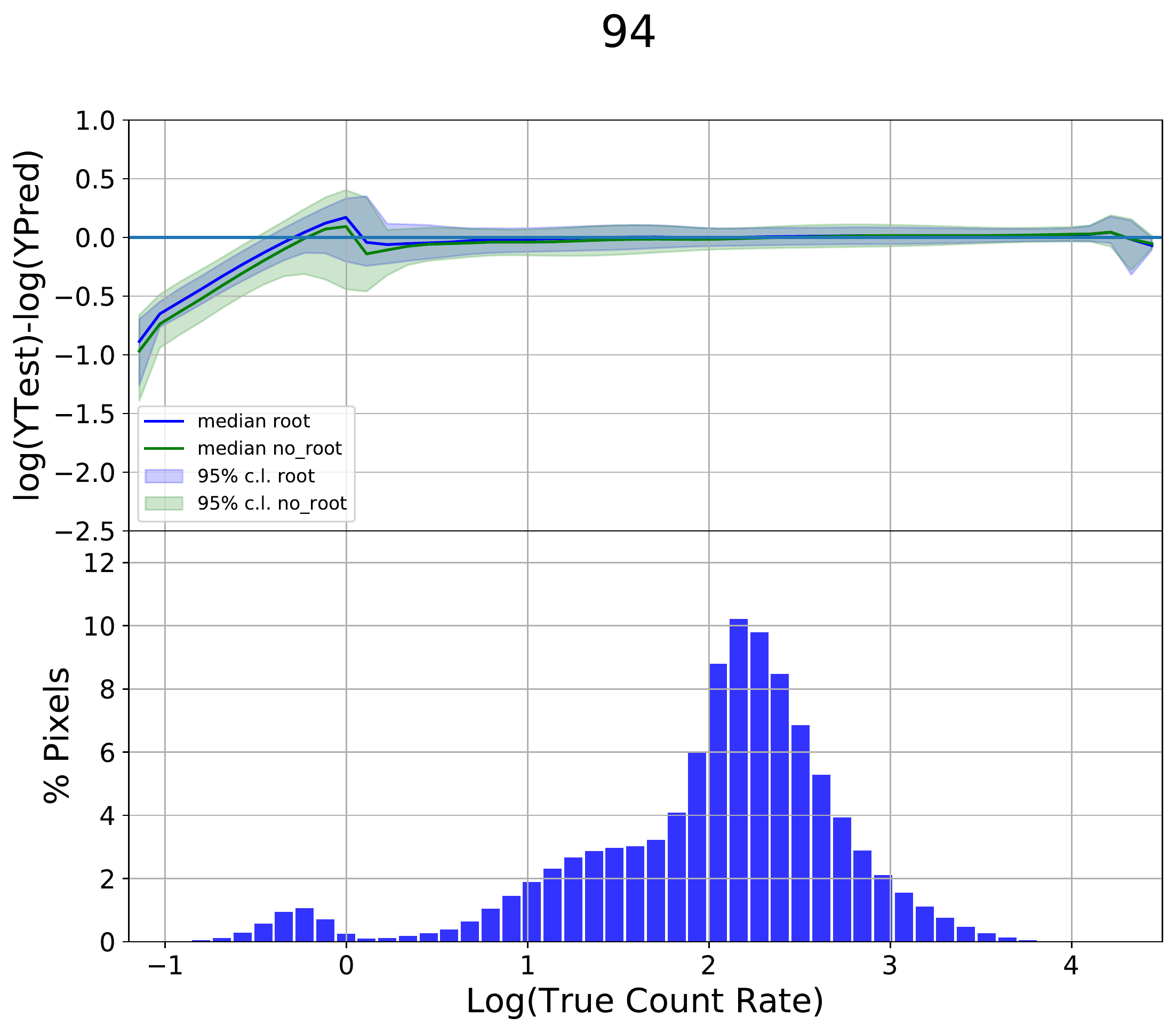}
  \caption{Predicted Intensity vs Real Intensity for each of the four channels, for all the pixels contained in the 667 images on the test set. From top to bottom: 211, 193, 171, 94 \AA{} channels. For each channel: the \textbf{top} plot shows the error on the predicted count rate as a function of the real count rate in $log10$. The error band represents the standard deviation, the line corresponds to the median. In green the standard U-net model, in blue the same architecture with square root scaling applied to the input images; the \textbf{bottom} plot shows the histogram of the pixel count rate distributions over the test set. The model performs well over three orders of magnitude but its accuracy degrades quickly in the extreme regions where fewer pixels are available.}
  \label{fig.combined_plots}
\end{figure}

 In Tab.~\ref{Tab-permutation} we explore the permutations of three input channels and one output channel and the effect of applying a root scaling transformation to the input images. In addition in Tab.~\ref{Tab-correlation} we show the correlation pixel by pixel for each of the permutations. We found that the same architecture produces similar reconstruction errors and correlation values over all the channels with a NMSE of about 0.01. We observe the similarity index of 94 \AA{} is worse of an order of magnitude with respect to the other channels, this can be explained by the fact SSIM is a not normalized metric and the average test value for this channel is higher than for the others (see Appendix, Tab.~\ref{tab:average_scaled_channels}). The results are remarkable, for example for 94 \AA{} the peak emission lies at a considerably higher temperature than the input channels \citep[see Fig.~1 of][]{Cheung:2015} that makes the reconstruction task a particularly challenging one. These results are in agreement with the results in \cite{salvatelli2019} and \cite{Lim_2021}. Please note that the values reported in Tab.1 of \cite{salvatelli2019} are not normalized. The squared-root scaling model shows roughly equivalent performance with the model with no squared-root applied to input data except for the channel 94~\AA{}.
 
 It is interesting to compare the results in Tab.~\ref{Tab-permutation} with those in Tab.~\ref{Tab-permutation-linear} where the same set of metrics are computed for the linear model. The DNN consistently improves by one order of magnitude over the linear model performance. This result clearly displays the value of using a DNN over a simpler model for the synthesis of the image. The comparison also demonstrates the strength of non-linearity between EUV channels and the fact it cannot be neglected for a meaningful reconstruction.
 
  In order to further evaluate the performance of both models, we calculate in Tab.~\ref{Tab-correlation} the average pixel-to-pixel Pearson correlation for pixels inside the solar disk for each channel combination. Agreeing with Tab.~\ref{Tab-permutation} results, the average pixel-to-pixel correlation shows both models have a remarkable performance where none of the channel combinations had a performance lower than 0.99. These results outperform all the channel combinations presented in \cite{Lim_2021}, which tries several combinations of EUV channels translations using the DL method ``Model B'' from \cite{Park_2019} and \cite{pix2pix}.
  
  Notably \cite{Lim_2021} did not report on other metrics we can use to compare the quality of the corresponding synthetic images. We demonstrate in the following analysis that the elevate visual quality of the images and the excellent pixel-to-pixel Pearson correlation values are not enough to guarantee the absence of artifacts which may impact the scientific utility of the synthetic images. This is illustrated in Fig.~\ref{fig.combined_plots}, Tab.~\ref{Tab-covariance211_flare} and Fig.~\ref{fig.covariance_plots}. Whether the discrepancies between real and synthetic images are sufficiently small to neglect clearly depends on the science case. For this reason, we argue that metrics such as covariance between real and synthetic image and accuracy by intensity should be standard metrics to be considered when reporting on models for the synthesis solar images.

 While useful to evaluate the overall performance of the algorithm, the aggregated metrics do not provide insights about the range of validity of the algorithm and the reasons behind its errors. Firstly, to understand how to possibly improve the model, and secondly, to clarify what could be a concrete use of the algorithm in future missions, it is helpful to evaluate the prediction uncertainty at different intensities. For all the permutations, in Fig.\ref{fig.combined_plots} we show the uncertainty on the predicted count rate (top) and the pixel distributions (bottom) as a function of the real count rate. These plots highlight three important factors:
 \begin{itemize}
     \item The algorithm does well over about three orders of magnitude of true count rate (intensity) and it largely increases its error when trying to predict the highest and lowest count rates. It means the global metrics would be much more favorable if removing these extreme pixels. This behavior also implies the algorithm could be used with confidence for applications that do not require accuracy on the most extreme values of count rates.
     \item The difficulty in predicting the pixels with the highest and the lowest count rate is not surprising if looking at the count rate distributions (histograms in Fig.\ref{fig.combined_plots}). The tails of the distributions, where the model's accuracy and uncertainty increase, are severely underrepresented in the distribution. This implies the image-to-image translation algorithm has not been trained or trained in a very limited way on pixels having these count rate values. This observation also provides a clear indication of which strategies can improve the algorithm performance, i.e., techniques to compensate the magnitude imbalance rather than larger architectures.
     \item Applying root scaling to the input images during the training tends to improve the results for low count rate pixels and reduces the uncertainty on the prediction. Some channels (193, 211 \AA{}) are more positively impacted than others by this change. This behavior is explained by the fact root scaling improves the sensitivity to small values during the training. We hypothesize that further exploration of different scaling strategies for the training can also be a way to extend the accuracy of the algorithm over more orders of magnitude.
 \end{itemize}
 
\begin{figure}
  \centering
  \includegraphics[width=1\linewidth]{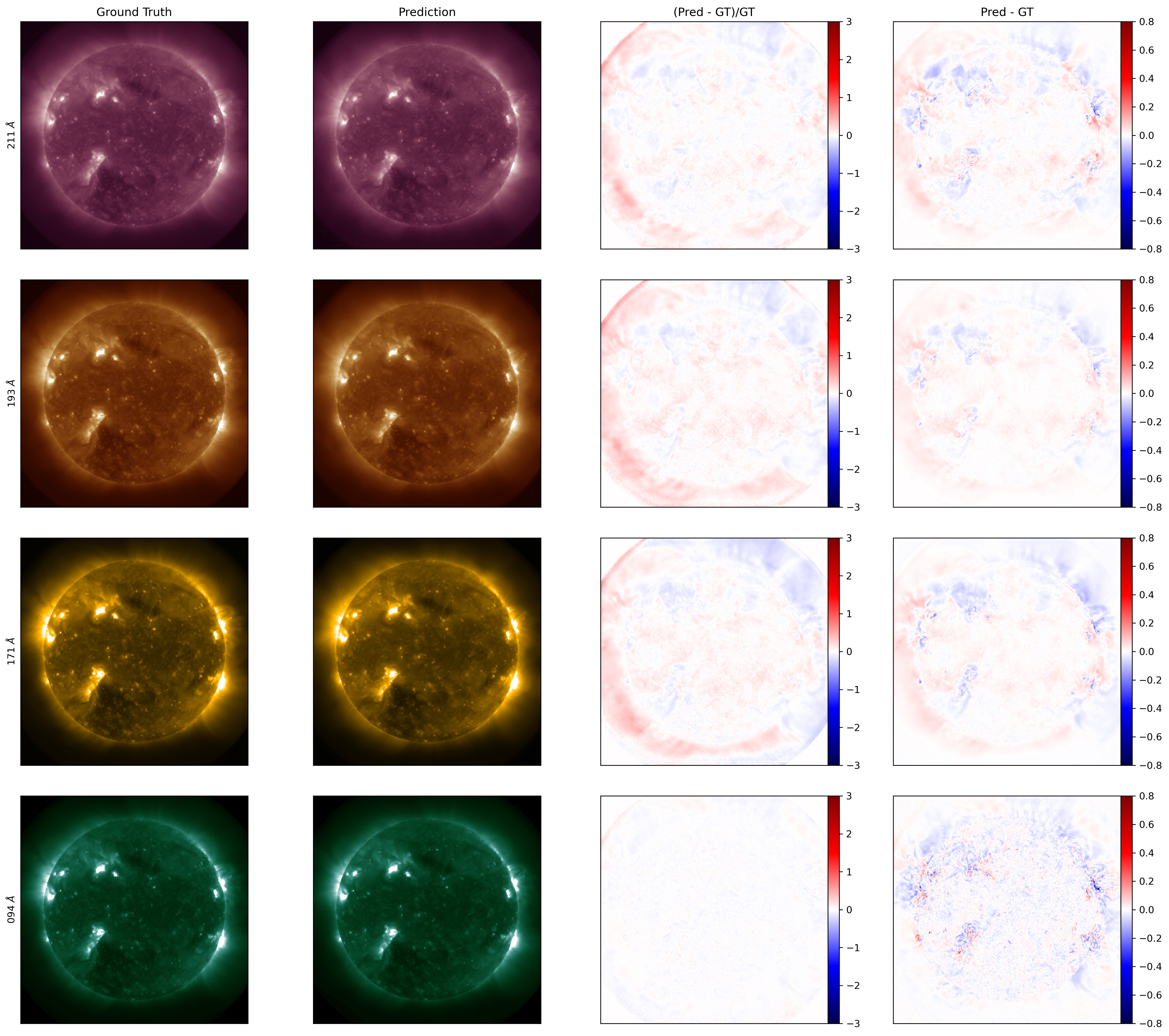}
  \caption{Real versus synthetic images on a quiet timestamp (2011-02-10 00:00:00 UT) when using model with root scaling. From left to right: real image, image synthesized by looking at the other 3 channels, residuals relative to the GT value} and difference between the two images. From top to bottom:  211, 193, 171, 94 \AA{} channels.
  \label{fig.gt_real_diff_quiet_root}
\end{figure}

\begin{figure}
  \centering
  \includegraphics[width=1\linewidth]{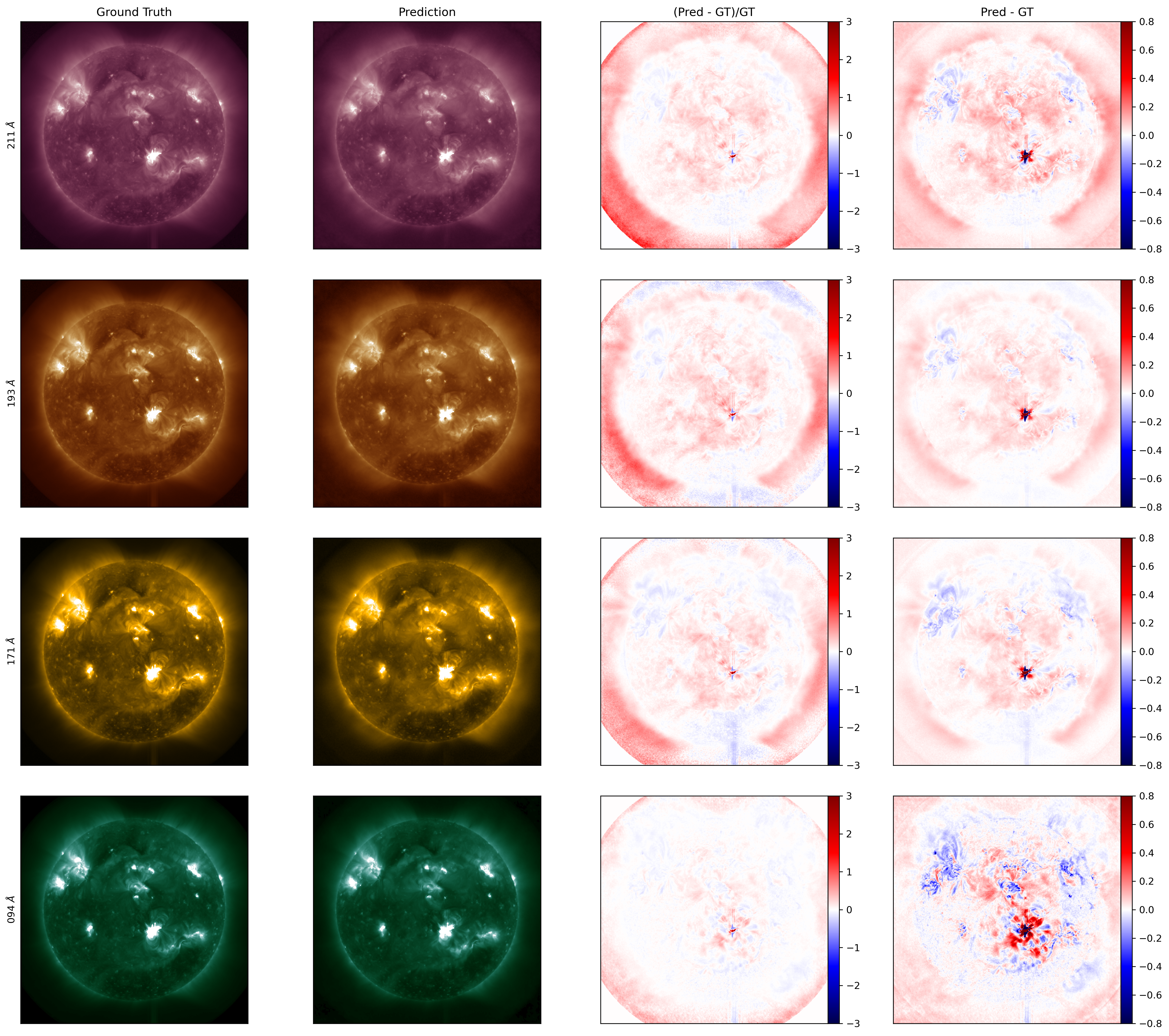}
  \caption{Real versus synthetic images during a flare (2011-02-15 02:00:00 UT) when using model with root scaling. From left to right: real image, image synthesized by looking at the other 3 channels, residuals relative to the GT value and} difference between the two images. From top to bottom: 211, 193, 171, 94 \AA{} channels.
  \label{fig.gt_real_diff_active_root}
\end{figure}
Examples of the resulting recovered images when adopting the DNN architecture described in Sec.~\ref{sec.meth} and a model with root scaled input, is given in Fig.~\ref{fig.gt_real_diff_quiet_root} and Fig.~\ref{fig.gt_real_diff_active_root}. The root scaling is reverted in the illustrated images. The first are example of reconstructions on a quiet day, where the Sun shows less activity, while the second are during the well known Valentine's Day flare. In these figures, the first column corresponds to the original images, while the second column corresponds to the ones generated by the DNN. Based on visual inspection, the synthetic image reproduces the morphology of coronal loops in the ground truth image for channels 211 and 171 \AA{}, and the prediction is instead a bit less realistic for 193 \AA{} for both quiet and active days. Clearly during the quiet day the all three channels have better performance than in the Valentine's day. It also interesting to observe that 94 \AA{} is the best performing channel during the quiet day, but the worst performing channel during the active day. This aligns to the results showed at Tab.\ref{Tab-permutation} and \ref{Tab-correlation}. It is unsurprising since the input AIA channels 94, 171 and 193 \AA{} channels have sensitivity to the plasma observed in the 211 \AA{} channel. This outperforms previous results in \cite{Park_2019}, where a conditional generative adversarial network (CGAN) had been trained to translate HMI magnetograms to AIA images.

In the third column of Fig.~\ref{fig.gt_real_diff_quiet_root} and Fig.~\ref{fig.gt_real_diff_active_root} we included the residuals relative to the real image and in the fourth column of the same figures we display the differences between the real and generated images. Dark blue and bright red correspond to the regions where the differences are the largest, and can be seen to be located where the active regions (shown as the brightest regions in the original and generated images) are.

\begin{figure}
  \centering
  \includegraphics[width=1\linewidth]{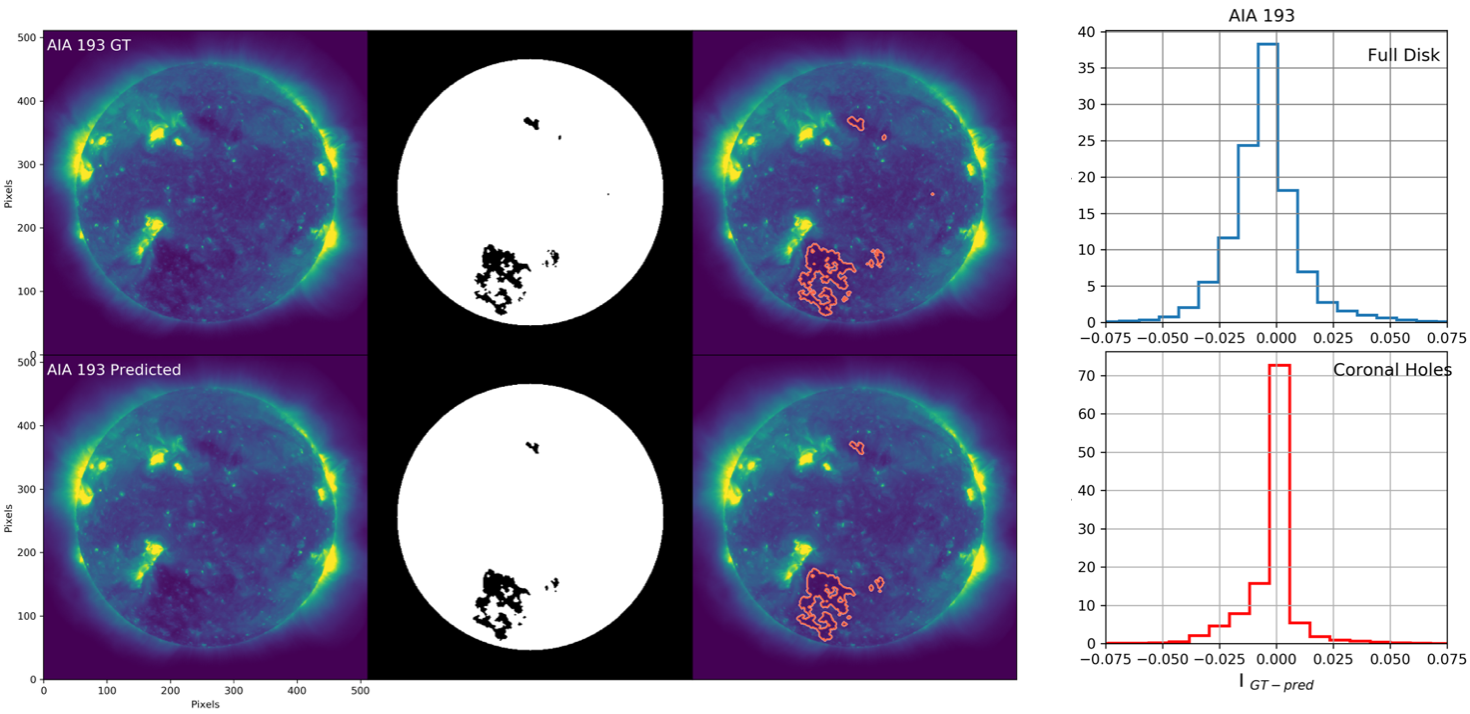}
  \caption{Coronal Holes for channel 193 \AA{}. On the left the segmentation mask obtained by thresholding , on the right histograms showing the difference between the ground truth and the predicted intensities (on a pixel-by-pixel basis) for both the pixels within the CHs and for the full-disk.}
  \label{fig.CH_analysis_193}
\end{figure}

\begin{figure}
  \includegraphics[trim=7 560 5 540, clip, width=1\linewidth]{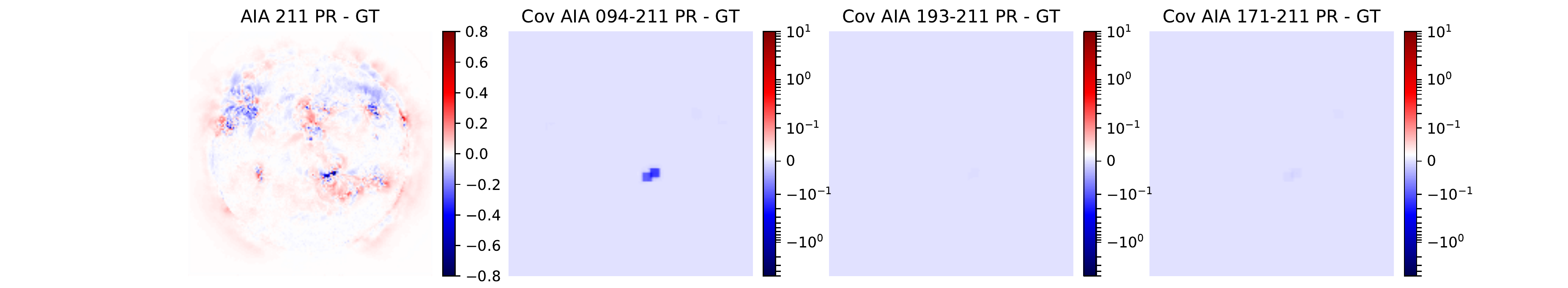}
  \newline
  \includegraphics[trim=7 560 5 530, clip, width=1\linewidth]{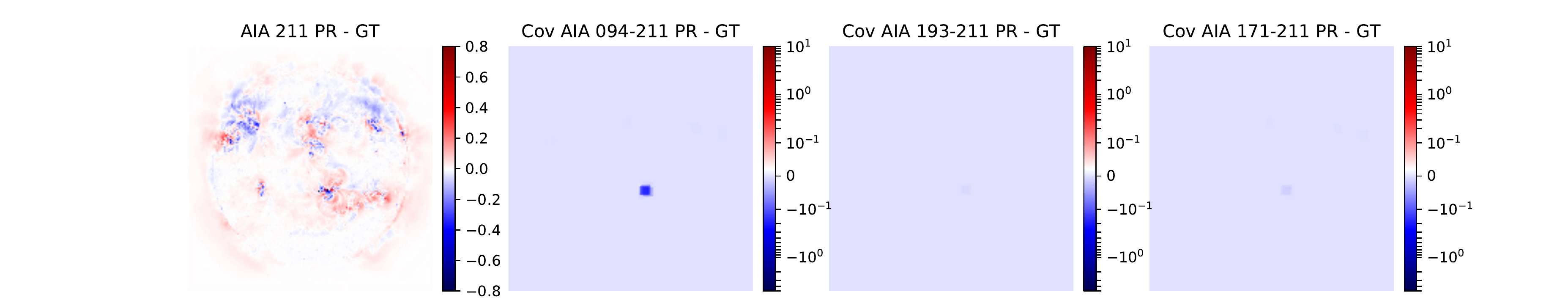}
  \newline
  \includegraphics[trim=7 560 5 530, clip, width=1\linewidth]{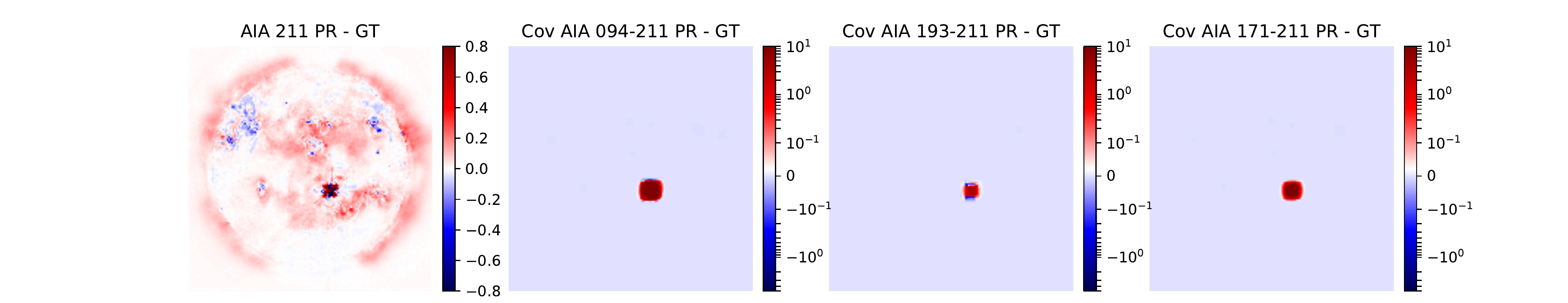}
  \newline
  \includegraphics[trim=7 560 5 530, clip, width=1\linewidth]{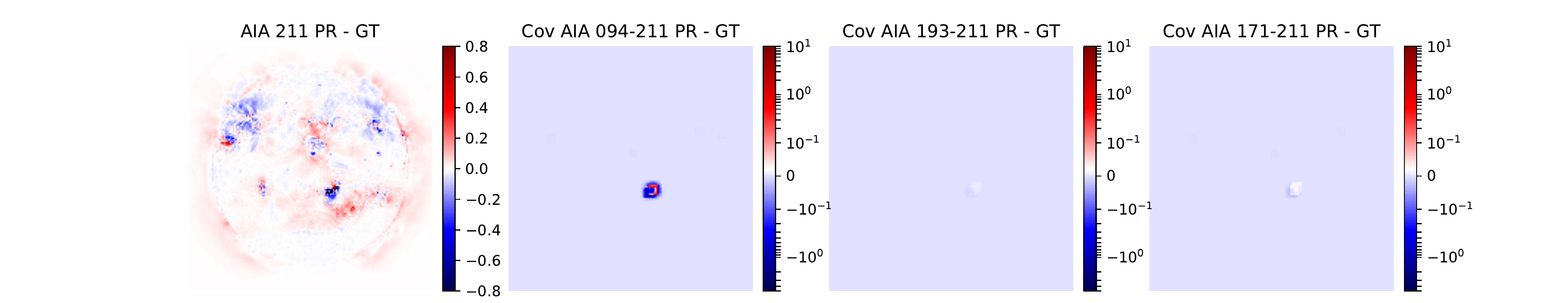}
  \newline
  \includegraphics[trim=7 560 5 530, clip, width=1\linewidth]{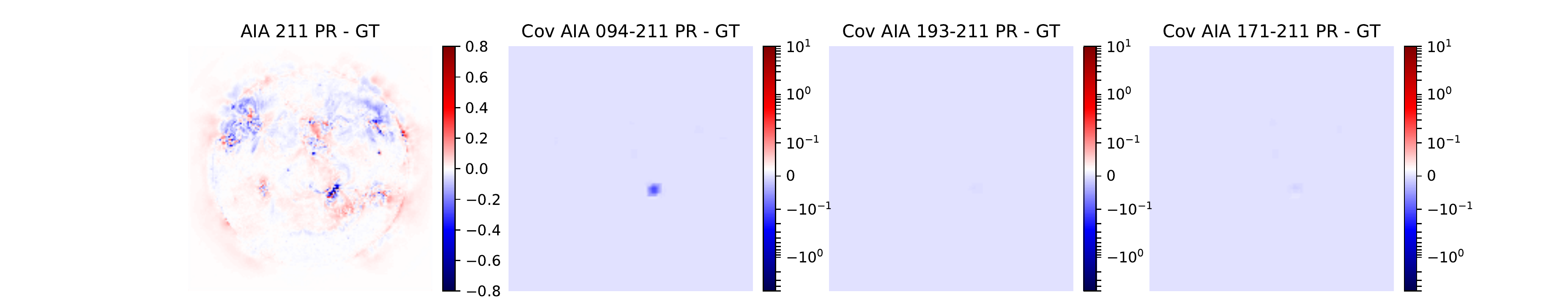}
  \caption{Reconstruction error on the covariance in correspondence of Valentine's Day flare. From left to right: Difference between the ground truth and the predicted images. Differences between the real and predicted covariance maps between 211 \AA{} - the predicted channel - and each of the input channels. From top to bottom: each row corresponds to a different timestamp at interval of 1h. The 3rd line is the closest to the time of the flare.}
  \label{fig.covariance_plots}
\end{figure}

Interestingly, the model well reconstructs coronal holes (CH) in both the active and quiet Sun cases described above, despite the low signal in these regions. This could be due to the fact that the physics of these regions is easier to model than active region coronal loops as the field lines are open and have relatively simpler configuration. A quantitative comparison between CH and full-disk is shown in Fig.~\ref{fig.CH_analysis_193} for channel 193 \AA{} (for the quiet Sun data represented in Fig.~\ref{fig.gt_real_diff_quiet_root}), where CHs are most distinctly visible due to their contrast. The segmentation mask identifies the CH regions based on the simple but robust adaptive intensity threshold technique \citep[similar to the technique employed in][]{2012SoPh..281..793R,2015SoPh..290.1355R}, and the histograms show the difference between the ground truth and the predicted intensities (on a pixel-by-pixel basis) for pixels both within the CH boundaries and the full-disk. It is to be noted that the segmentation mask is constructed for both the predicted and ground-truth images independently using the same intensity threshold criterion. Clearly, the predicted AIA intensities are well constrained not just over the full disk but also on the relatively quieter CH areas.

\begin{table}
\centering
\begin{tabular}{l|c|cccc}
\toprule
     Timestamp &   Channel &  True Cov &  Pred Cov &    Diff &  \%Diff \\
\midrule
 2011-2-15-0-0 &     94 &     0.278 &     0.256 &   0.022 &    7.9 \\
 2011-2-15-1-0 &     94 &     0.262 &     0.246 &   0.016 &    5.9 \\
 2011-2-15-2-0 &     94 &     13.9 &      92.3 & -78.5 &      -565 \\
 2011-2-15-3-0 &     94 &     1.69 &     1.54 &   0.150 &    8.9 \\
 2011-2-15-4-0 &     94 &     0.392 &     0.375 &   0.017 &    4.4 \\ \\
 2011-2-15-0-0 &     171 &     0.117 &     0.115 &   0.002 &    2.1 \\
 2011-2-15-1-0 &     171 &     0.114 &     0.112 &   0.002 &    1.9 \\
 2011-2-15-2-0 &     171 &     1.29 &    13.1 & -11.8 & -913 \\
 2011-2-15-3-0 &     171 &     0.186 &     0.178 &   0.008 &    4.3 \\
 2011-2-15-4-0 &     171 &     0.139 &     0.136 &   0.003 &    2.3 \\ \\
 2011-2-15-0-0 &     193 &     0.048 &     0.047 &   0.001 &    1.4 \\
 2011-2-15-1-0 &     193 &     0.047 &     0.047 &   0.001 &    1.3 \\
 2011-2-15-2-0 &     193 &     0.191 &     0.605 &  -0.414 & -216 \\
 2011-2-15-3-0 &     193 &     0.065 &     0.063 &   0.003 &    4.0 \\
 2011-2-15-4-0 &     193 &     0.055 &     0.054 &   0.001 &    2.1 \\
\bottomrule
\end{tabular}
\caption{Errors in reconstructing the covariance between 211 \AA~ and the other 3 channels when using the synthetically produced image for 211 \AA{} in correspondence of a highly energetic event (Valentine's Day flare on 2011-2-15:1:50:00 UT). Interestingly the reconstructed covariance has a much higher error than what seen in a quiet period, cf. Tab.~\ref{Tab-covariance211_quiet}, at least 1h before the flare has been detected.}
\label{Tab-covariance211_flare}
\end{table}

\begin{table}
\centering
\begin{tabular}{l|c|cccc}
\toprule
     Timestamp &   Channel &  True Cov &  Pred Cov &    Diff &  \%Diff \\
\midrule
 2011-2-13-0-0 &     94 &    0.1506 &    0.1504 &  0.0002 &    0.1 \\
 2011-2-13-1-0 &     94 &    0.1672 &    0.1654 &  0.0018 &    1.1 \\
 2011-2-13-2-0 &     94 &    0.1601 &    0.1588 &  0.0013 &    0.8 \\
 2011-2-13-3-0 &     94 &    0.1713 &    0.1718 & -0.0004 &   -0.3 \\
 2011-2-13-4-0 &     94 &    0.1652 &    0.1650 &  0.0002 &    0.1 \\ \\
 2011-2-13-0-0 &     171 &    0.1213 &    0.1210 &  0.0002 &    0.2 \\
 2011-2-13-1-0 &     171 &    0.1261 &    0.1254 &  0.0007 &    0.5 \\
 2011-2-13-2-0 &     171 &    0.1227 &    0.1223 &  0.0004 &    0.3 \\
 2011-2-13-3-0 &     171 &    0.1241 &    0.1244 & -0.0002 &   -0.2 \\
 2011-2-13-4-0 &     171 &    0.1226 &    0.1219 &  0.0007 &    0.6 \\ \\
 2011-2-13-0-0 &     193 &    0.0449 &    0.0448 &  0.0000 &    0.1 \\
 2011-2-13-1-0 &     193 &    0.0470 &    0.0468 &  0.0002 &    0.4 \\
 2011-2-13-2-0 &     193 &    0.0439 &    0.0439 & -0.0000 &   -0.1 \\
 2011-2-13-3-0 &     193 &    0.0465 &    0.0468 & -0.0003 &   -0.7 \\
 2011-2-13-4-0 &     193 &    0.0471 &    0.0470 &  0.0001 &    0.2 \\
\bottomrule
\end{tabular}
\caption{Errors in reconstructing the covariance between 211 \AA{} and the other 3 channels when using the synthetically produced image for 211 \AA{} in correspondence of a quiet period few days before Valentine's Day flare. The percentage difference is below 1\% for all the channels.}
\label{Tab-covariance211_quiet}
\end{table}
In Tab.~\ref{Tab-covariance211_flare} and Tab.~\ref{Tab-covariance211_quiet} we report the reconstruction error on the covariance between channels, over four hours, for the case 94, 171, 193 \AA{} to 211 \AA{} in correspondence of a flare and on a normally quiet day. Not surprisingly, in light of the results above, the reconstructed covariance has great accuracy (less than 1\% of error) on a quiet day but its error increases in several orders of magnitude in correspondence of the extreme event. The results reported in Tab.~\ref{Tab-covariance211_flare} and Tab.~\ref{Tab-covariance211_quiet} are obtained using the model without square root scaling, the most sensitive to extreme values. They should therefore be interpreted as an upper bound on the error that a similar image translation would have. 
With the aim of better understanding the source of error, in addition to the standard covariance, we compute a covariance map with spatial mean on a rolling squared window of $20\times20$ pixels, see Eq.~\ref{eq.patch_covariance} for definition. The resulting covariance map in correspondence of a flare is shown in Fig.~\ref{fig.covariance_plots}. The map clearly shows the error of the model is localised in the area of the flare and it does not affect the rest of the map, in agreement with the localized reconstruction error shown in Fig.~\ref{fig.gt_real_diff_active_root}. This result confirms the results of the ``virtual telescope" would be accurate for most of the pixels, also in presence of an extremely energetic event, but for the specific area where the event happens. Similar results hold for the covariance in other channel permutations. 

Incidentally, the above covariance result suggests an increase in its reconstruction error could also be used as a method for early detection of flares as the error starts to increase before the actual flare's event. Variations in reconstruction errors are commonly used in machine learning as anomaly detection methods (e.g. ~\cite{An2015VariationalAB, anomaly_detection}. While directly detecting an increase in the data count could be found to be more effective, the sensitivity to non-linearity of the reconstruction task could produce a stronger or complementary signal that we think is interesting to consider in future work.

\section{Concluding remarks}
\label{section:conlusions}
In this study, we analyzed the performance of an image-to-image translation DNN model in accurately reconstructing extreme ultra-violet images from a solar telescope, focusing on the permutations of four channels. We found that the reconstruction error is extremely accurate over three orders of magnitude in pixel intensity (count rate) and it rapidly increases when considering extremely low and high range of intensities. This behavior is explained by the pixel count rate distribution in the training set, the rarer the value the more difficult for the DNN to provide an accurate prediction. Similarly, when looking at the reconstruction error on the covariance at different times, we found the model can synthetically predict the covariance with less than 1\% of error on quiet days but its performance is severely affected in correspondence of flares, in the active regions. 

The results show that a virtual telescope would produce accurate estimations on a range of intensities but, if built following the methodology here described, would not be able to accurately reproduce extremely energetic events like flares. How and in which limit the reconstruction error for such specific events could be improved is an area of research that we leave for future work. The rareness of flare events poses a challenge in training machine learning algorithms to accurately reproduce such events. Based on the results above, we think adopting oversampling techniques and different scaling strategies would improve at least in some measure the performance. To overcome this challenge, other strategies like automatic detection of anomalies could also be adopted in combination with image-to-image translation, in the design of a virtual solar telescope.

In this paper, we did not explore the dependence of model performance from spatial resolution. In principle smaller subpixel scales could have information that improve the global performance of image synthesis and we think this is an important question to be addressed in future work. Importantly, we expect the deterioration of the synthetic accuracy for rare events to happen regardless of the adopted scale because it is caused by the scarcity of examples for training.

\hspace{2cm}

\textbf{Acknowledgments}
 This project has been initiated during the 2019 NASA Frontier Development Lab (FDL) program, a public/private partnership between NASA, SETI and industry partners including Lockheed Martin, IBM, Google Cloud, NVIDIA Corporation and Intel. We thank all our FDL mentors for useful discussion in the early stage of the projec, as well as the SETI Institute for their support during the program and beyond. L.F.G.S acknowledges support from NASA under Grant No. 80NSSC20K1580. M.C.M.C. and M.J. acknowledge support from NASA’s SDO/AIA (NNG04EA00C) contract to the LMSAL. S.B. gratefully acknowledges support from NASA contracts NNG09FA40C (IRIS) and 80NSSC20K1272. We thank the NASA’s Living With a Star Program, which SDO is part of, with AIA, and HMI instruments on-board.

Software: We acknowledge for CUDA processing cuDNN \citep{cudnn}, for data analysis and processing we used Numpy \citep{numpy}, Pandas \citep{pandas}, SciPy \citep{scipy} and scikit-learn\citep{scikit-learn}. All plots were done using Matplotlib \citep{matplotlib}.

\clearpage
\appendix
\section{Scaling units for each AIA channel}
\label{section:appendix_average}
\begin{table}[htb!]
  \centering
  \begin{tabular}{cc}
    \toprule
     AIA channel (\AA) &  Scaling unit [DN/s/pixel] \\
     \midrule
       94 &   10  \\
      171 & 2000  \\
      193 & 3000  \\
      211 & 1000  \\
      \bottomrule
  \end{tabular}
  \caption{Table of  AIA channel scaling units.}
  \label{tab:average_channels}
\end{table}
\begin{table}[htb!]
  \centering
  \begin{tabular}{cc}
    \toprule
     AIA channel (\AA) &  $\overline{Y_{test}}$ \\
     \midrule
       94 &   26  \\
      171 & 0.13  \\
      193 & 0.087  \\
      211 &  0.26  \\
      \bottomrule
  \end{tabular}
  \caption{Table of average values over the test set after scaling by channel}
  \label{tab:average_scaled_channels}
\end{table}

\section{Code description}
\label{section:App_A_codebase}
In this appendix we describe the modular software used to produce the analysis and made freely available online on GitHub under GPL licence. Users are invited to consult the code documentation for additional detail.

\begin{itemize}
    \item \textit{src/sdo} - contains all the modules required to run the pipeline plus additional functionalities that can be used as standalone library to interact with the SDO-ML dataset v1.
    \item \textit{config} - contains some configuration templates.
    \item \textit{scripts} - contains some analysis scripts specific to the paper, they can be used to reproduce the results.
    \item \textit{notebooks} - contains some notebooks specific to the paper that can be used to reproduce some of the plots in the paper and some examples to show how to use some functionalities (e.g. how to use the dataloader to load timestamps of interest).
\end{itemize}

The most relevant modules under \textit{src} are:
\begin{itemize}
    \item \textit{src/sdo/datasets/sdo\_dataset.py} this module contains the \textit{SDO\_Dataset} class, a custom Dataset class compatible with \textit{torch.utils.data.DataLoader}. It can be used to flexibly load a train or test dataset from the SDO local folder. Data can be selected according to the 3 criteria:
    \SubItem{asking for a specific range of years and a specific frequency in months, days, hours, minutes}
    \SubItem{passing a file that contains all the timestamps of interest}
    \SubItem{passing two timestamps ranges and a desired step}
    
    This class assumes a pre-computed inventory of the SDO dataset exists.
    
    \item \textit{src/sdo/pipelines/virtual\_telescope\_pipeline.py} this module contains the \textit{VirtualTelescopePipeline} class, the class that contains all the training and test logic of the modeling approach. This class also handles the metrics logging and the files saving. Beyond being used for reproducing the results of this work, this class can be used as example of how to integrate the dataloader above with other PyTorch models for a different set of experiments.
    
    \item \textit{src/sdo/parse\_args.py} this module contains the description of all the parameters that can be passed as input to the pipeline and their default values.
\end{itemize}

\section{Additional Figures}
In this appendix we report some additional results not included in the main text.
\begin{figure}[htb!]
  \centering
  \includegraphics[width=1\linewidth]{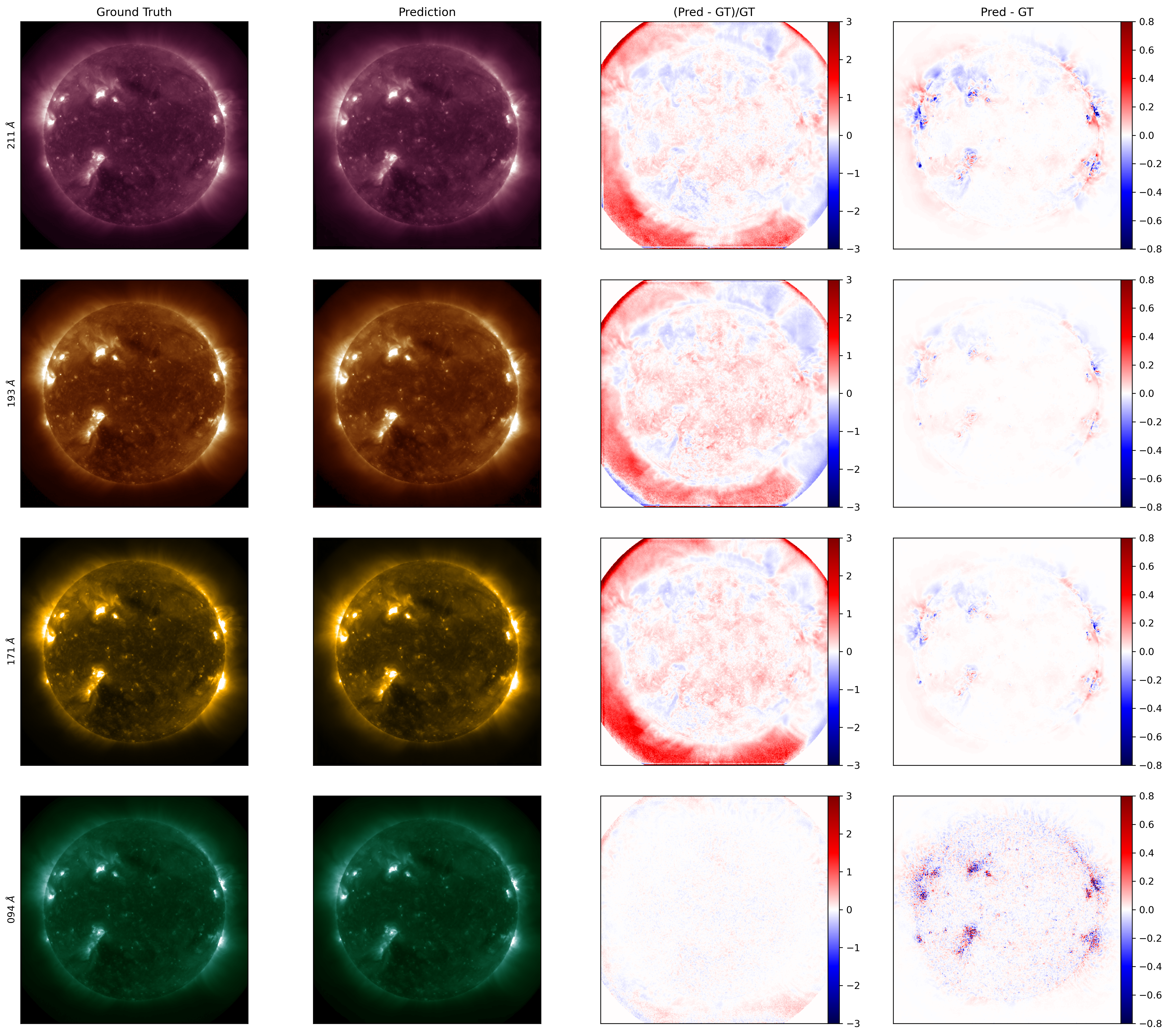}
  \caption{Real versus synthetic images on a quiet timestamp (2011-02-10 00:00:00) when using model without root scaling. From left to right: real image, image synthesized by looking at the other 3 channels, residuals relative to the GT value and difference between the two images. From top to bottom 211, 193, 171, 94 \AA{} channels.}
  \label{fig.gt_real_diff_quiet_nr}
\end{figure}
\begin{figure}[htb!]
  \centering
  \includegraphics[width=1\linewidth]{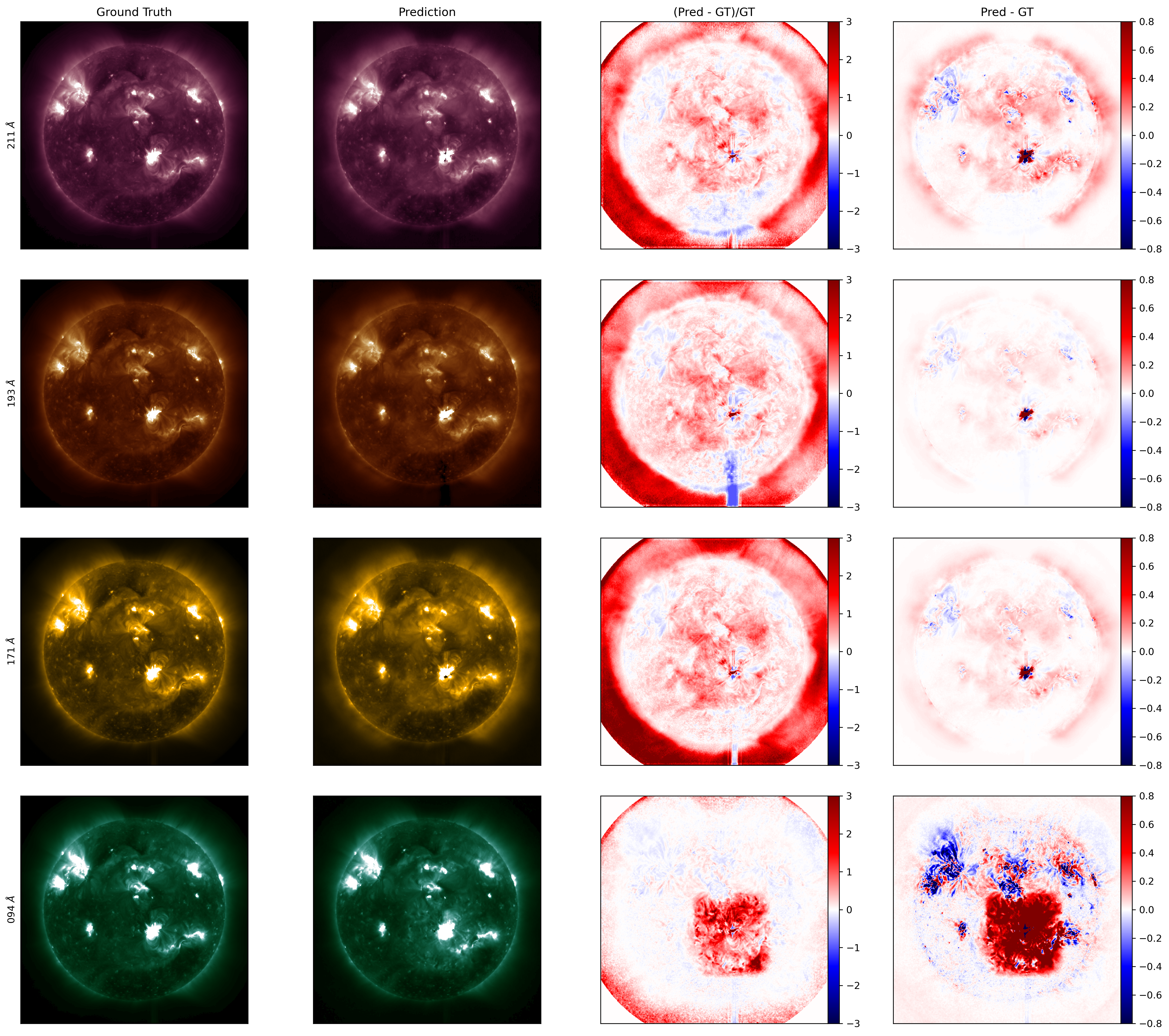}
  \caption{Real versus synthetic images during a flare (2011-02-15 02:00:00)  when using model without root scaling. From left to right: real image, image synthesized by looking at the other 3 channels, residuals relative to the GT value and difference between the two images. From top to bottom 211, 193, 171, 94 \AA{} channels.\color{red}}
  \label{fig.gt_real_diff_active}
\end{figure}

\clearpage

\bibliography{virtual_telescope}{}
\bibliographystyle{aasjournal}



\end{document}